\documentclass[journal,,draftclsnofoot letterpaper, onecolumn]{IEEEtran}

\usepackage{cite}
\usepackage{graphicx}
\usepackage{subfigure}
\usepackage{amsmath}
\usepackage{amssymb}
\usepackage{bm}
\usepackage{epsfig}
\usepackage{times}
\usepackage{color}
\usepackage{url}
\usepackage{array}
\usepackage{multirow}
\usepackage{rotating}
\usepackage{extarrows}
\usepackage{amsthm}
\usepackage{setspace}
\usepackage{tikz}
\usetikzlibrary{arrows,shapes}
\usepackage{epstopdf}

\begin{document}

\title{Beam-forming for Secure Communication in Amplify-and-Forward Networks: An SNR based approach}%
\author{Siddhartha Sarma, Samar Agnihotri, and Joy Kuri%
\thanks{S. Sarma and J. Kuri are with the Department of Electronic Systems Engineering, Indian Institute of Science, Bangalore, Karnataka - 560012, India (e-mail: \{siddharth, kuri\}@dese.iisc.ernet.in).}
\thanks{S. Agnihotri is with the School of Computing and Electrical Engineering, Indian Institute of Technology Mandi, Mandi, Himachal Pradesh - 175001, India (e-mail: samar@iitmandi.ac.in).}
}

\maketitle

\begin{abstract}
The problem of secure communication in Amplify-and-Forward (AF) relay networks with multiple eavesdroppers is considered. Assuming that a receiver (destination or eavesdropper) can decode a message only if the received SNR is above a predefined threshold, we introduce SNR based optimization formulations to calculate optimal scaling factors for relay nodes in two scenarios. In the first scenario, we maximize the achievable rate at the legitimate destination, subject to the condition that the received SNR at each eavesdropper is below the target threshold. Due to the non-convex nature of the objective function and eavesdroppers' constraints, we transform variables and obtain a \textit{Quadratically Constrained Quadratic Program} (QCQP) with convex constraints, which can be solved efficiently. When the constraints are not convex, we consider a \textit{Semi-definite relaxation} (SDR). In the second scenario, we minimize the total power consumed by all relay nodes, subject to the condition that the received SNR at the legitimate destination is above the threshold and at every eavesdropper, it is below the corresponding threshold. We propose a \textit{semi-definite relaxation} of the problem in this scenario and also provide an analytical lower bound. 
\end{abstract}

\section{Introduction}
In wireless networks, generally security measures are implemented in the upper layers of protocol stack using cryptographic methods. However, current advances in computation technology pose threats for such systems, prompting researchers to explore alternatives like \textit{physical layer security} \cite{bloch2011physical}. In ad-hoc wireless networks, where relay nodes are constrained by hardware or available energy resources, physical-layer based schemes can indeed provide effective solutions. 

In the physical-layer security literature, most results are centered around an information-theoretic measure - the \textit{secrecy capacity} \cite{wyner}. The secrecy capacity is the maximum rate of information transfer from the source to the destination in the presence of eavesdropper(s). 
However, even for many simple networks, calculating secrecy capacity for a given eavesdropper model is a tedious task as it involves solving fractional optimization problems, which are known to be NP-hard in general \cite{schaible}. For amplify-and-forward (AF) relay networks, the authors in \cite{dong} evaluated secrecy capacity for a single eavesdropper, and the zero-forcing secrecy rate for multiple eavesdroppers under a total relay power constraint. In \cite{zhang10} and \cite{yang2013cooperative}, techniques are devised for computing the maximum AF secrecy rate with a single and multiple eavesdroppers under both individual and total relay power constraints, respectively. However, our work differs from them as we have considered an SNR based model for secure communication.
%
The motivation for our approach arises from the results in
\cite{Leung}, where it is shown that the secrecy capacity depends on the source-destination and the source-eavesdropper channel capacities, which in turn are functions of the respective signal-to-noise ratios (SNR).

In practical receivers, the correct decoding of the received signals is dependent on the received SNR \cite{proakisdigital}. If the received SNR is below a certain predefined threshold, then for all practical purposes, we can assume that the receiver cannot extract any information from the received signal. Following this argument, we consider a model where every eavesdropper has limited decoding capability, determined by its SNR threshold. Therefore, to ensure secure communication between the source and the destination, we strive to bring down the received SNR at each eavesdropper below its target SNR threshold. This approach is a generalization of the zero-forcing approaches for secure communication \cite{dong}. Concurrently, we also strive to improve the SNR at the legitimate destination. This leads us to our first scenario, where we maximize the achievable rate at the legitimate destination, subject to the condition that the received SNR at each eavesdropper is below its target threshold. We address this scenario in Section \ref{SNR}. Though a similar problem was considered in \cite{yang2013cooperative}, but unlike them we are not interested in evaluating optimal secrecy rate. Infact, as each eavesdropper can have a different threshold, therefore the optimal solution of our problem will differ from the solution of \textit{Secrecy Rate Maximization} (SRM) problem considered in \cite{yang2013cooperative}. Also, we propose a noble transformation for this problem to obtain a \textit{Quadratic Constrained Quadratic Program} (QCQP) which attains global solution under certain criteria.


The second scenario that we consider is motivated by wireless adhoc networks, where the intermediate relay nodes may be interested in minimizing total power consumption while ensuring secure communication between the source and the destination.
 In such scenarios, we would like to minimize the total power consumption of relay nodes, subject to the condition that the received SNR at the legitimate destination is above a predefined threshold and at every eavesdropper it is below the respective predefined threshold.
This formulation is described in Section \ref{POW}. 

A similar SNR based model is also considered in \cite{sarma}. However, unlike the decode-and-forward (DF) relaying used in that work, we employ AF relaying as it is one of the simplest relaying schemes and even in low-SNR regimes, it can be a capacity achieving relay strategy in some scenarios \cite{107gomadamJafar}.


\section{System Model}
\label{sec:model}
The system consists of a single source $(S)$, a single destination $(D)$, $M$ relay nodes $(i \in \mathcal{M}=\{1,2,\cdots,M\})$ and $K$ eavesdroppers $(k \in \mathcal{K}=\{1,2,\cdots,K\})$ as shown in Figure \ref{fig:model}. The channel gain from a node $p$ to a node $q$ is denoted by a complex constant $h_{p,q}$, where $p\in \{S\}\bigcup \mathcal{M}$ and $q \in \{D\} \bigcup \mathcal{M} \bigcup \mathcal{K}$. We assume the availability of complete channel state information 
(CSI), \textit{i.e.}, all $h_{p,q}$ values are known throughout the network \cite{dong}, \cite{sarma}. Considering  discrete time instants and neglecting propagation delays, the signal received at each relay node due to the source can be expressed as:
\begin{equation}\label{eq:rel}
y_i[n]= h_{s,i}x_s[n] + z_i[n]
\end{equation} 
where the source signal $x_s[n]$ at time instant $n, -\infty < n < \infty$, are i.i.d complex Gaussian random variables with zero mean and variance $P_s$. 
Further, $x_s[n]$ is assumed to be independent of $z_i[n]$, the complex white Gaussian noise $(\sim \mathcal{CN}(0,\sigma^2))$ at the $i^{th}$ relay node. Each relay node $i$ scales the received signal from the source by $\beta_i \in \mathbb{C}$ before transmitting: 
\begin{figure}[!t]
\centering
\begin{tikzpicture}[scale=1.25]
\tikzstyle{every node}=[draw,shape=circle,minimum size=0.9cm,style=thick]
\node (v0) at (180:2.5) {$S$};
\node (v1) at (90:1.5) {$1$};
\node (v2) at (90:0.5) {$2$};
\node (v3) at (270:1) {$M$};
\node (v4) at (25:3) {$D$};
\node (v5) at (5:2.85) {$E_1$};
\node (v6) at (340:3.05) {$E_K$};
\draw[style=thick,->] (v0) -- (v1);
\draw[style=thick,->] (v0) -- (v2);
\draw[style=thick,->] (v0) -- (v3);
\draw[style=thick,->] (v1) -- (v4);
\draw[style=thick,->] (v2) -- (v4);
\draw[style=thick,->] (v3) -- (v4);
\draw[dashed,->](v1) -- (v5);
\draw[dashed,->](v2) -- (v5);
\draw[dashed,->](v3) -- (v5);
\draw[dashed,->](v1) -- (v6);
\draw[dashed,->](v2) -- (v6);
\draw[dashed,->](v3) -- (v6);
\draw[loosely dotted,line width=1](0,0.10) -- (0,-.6);
\draw[loosely dotted,line width=1](2.9,-0.15) -- (2.9,-.6);
\draw[dashed,,rounded corners] (2.4,0.65) rectangle (3.3,-1.5);
\end{tikzpicture}
\caption{Simple AF network with multiple ($K > 1$) eavesdroppers.}
\label{fig:model}
\end{figure}
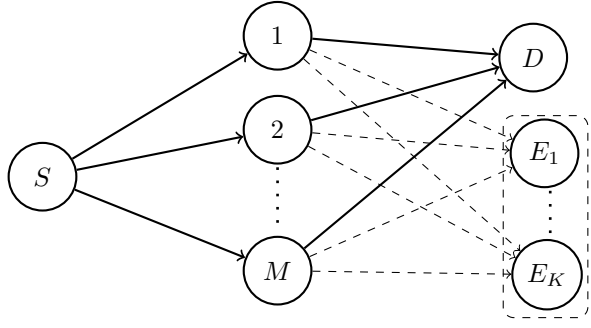
\begin{equation}
\label{eq:beta}
x_i[n+1]=\beta_i y_i[n],\;|\beta_i|^2 \le \beta_{i,max}^2= \frac{P_i}{|h_{s,i}|^2 P_s + \sigma^2}
\end{equation}
where 
$\mathbf{E}[|x_i[n]|^2] \le P_i,\;\forall i \in \mathcal{M}$. Therefore, the received signal at the destination or an eavesdropper can be expressed as:
\begin{align}
y_l[n] = & \sum\limits_{i=1}^{M}h_{i,l}x_i[n]+z_l[n],\;l \in \{D\}\bigcup \mathcal{K} \label{eq:dest}
\end{align}
Here $z_l[n] \sim \mathcal{CN}(0,\sigma^2)$. We assume that $z_i$ are independent of the source symbol and of each other. For the layered relay network shown in Figure \ref{fig:model}, 
the received signals at the destination and the eavesdroppers are free of intersymbol interference as they traverse paths with the same respective delays. Therefore, omitting the time index and using \eqref{eq:rel}, \eqref{eq:beta} and \eqref{eq:dest} we can write:
\begin{equation}\label{eq:rcvsgnl}
y_l=\sum\limits_{i=1}^{M}h_{s,i}\beta_ih_{i,l}x_s + \sum\limits_{i=1}^{M}\beta_ih_{i,l}z_i + z_l
\end{equation}
From \eqref{eq:rcvsgnl}, we can write the SNR at the destination or an eavesdropper as:
\begin{equation*}
SNR_l=  \frac{\left|\sum\limits_{i=1}^{M} h_{s,i}\beta_ih_{i,l}\right|^2}{1 + \sum\limits_{i=1}^{M}|\beta_i h_{i,l}|^2}\frac{P_s}{\sigma^2}=\frac{|\mathbf{h_{s,l}}^\dagger\bm{\beta}|^2}{1+\bm{\beta}^\dagger \mathbf{D}_l\bm{\beta}}\frac{P_s}{\sigma^2}
\end{equation*} 
where $\mathbf{h_{s,l}}$ is an $M \times 1$ column vector $\mathbf{h_{s,l}}=[h_{s,1}h_{1,l},h_{s,2}h_{2,l},\ldots,h_{s,M}h_{M,l}]^\dagger $ and $\mathbf{D}_l$ is a $M \times M$ diagonal matrix \textit{i.e.}, $\mathbf{D}_l = diag(|h_{1,l}|^2, \ldots, |h_{M,l}|^2), l \in \{D\} \bigcup \mathcal{K}$.
  

\section{Receiver SNR Maximization}
\label{SNR}
In the first scenario, we consider the problem of maximizing the secrecy rate achievable with the AF-relays when the received SNR at the eavesdroppers is below their respective predefined thresholds. However, given the logarithmic dependence of the achievable information rate on the received SNR ($I(P_s) \propto \log(1+SNR)$) and the monotonically increasing nature of the $\log(\cdot)$, we have:
      \begin{subequations}\label{prob2}
      \begin{align}
       \text{\textbf{P1:}}\;\max_{\bm{\beta}} \; & \frac{\left| \sum_{i=1}^{M} h_{s,i}\beta_ih_{i,d}\right| ^2}{1 + \sum_{i=1}^{M}|\beta_i h_{i,d}|^2}\frac{P_s}{\sigma^2} \\
      \text{such that}  \; & \frac{\left| \sum_{i=1}^{M} h_{s,i} \beta_i h_{i,k}\right| ^2}{1 + \sum_{i=1}^{M}|\beta_ih_{i,k}|^2}\frac{P_s}{\sigma^2} \le \gamma_k, \; k \in \mathcal{K}\\
       		                &  |\beta_i|^2 \le \beta_{max,i}^2, i \in \mathcal{M}
      \end{align}
      \end{subequations}
      Here, $\gamma_k$ is a real number and represents the predefined threshold for the $k^{th}$ eavesdropper. This problem is inherently non-convex. We consider the following transformation of the variable $\beta_i$.
       \begin{align*}
       \omega_i=\beta_ih_{i,d}&,\;\text{and } u_i=\frac{\omega_i}{\sqrt{1+\bm{\omega}^\dagger\bm{\omega}}}
       \end{align*}
       If we consider the vector variables $\mathbf{u}=[u_1,u_2,\cdots,u_M]^\mathbf{t}$ and $\bm{\omega}=[\omega_1,\omega_2,\cdots,\omega_M]^\mathbf{t}$, then we can write:
       \begin{align*}
        \mathbf{u}=\frac{\bm{\omega}}{\sqrt{1+\bm{\omega}^\dagger\bm{\omega}}} \Leftrightarrow \bm{\omega}=\frac{\mathbf{u}}{\sqrt{1-\mathbf{u^\dagger u}}}
       \end{align*}
       Also, we define $\bm{\rho}_k = [\rho_{1,k}, \ldots, \rho_{M,k}]$, where $\rho_{i,k}=\frac{h_{i,k}}{h_{i,d}}$.

In terms of these new variables and parameters, the objective function and constraints are expressed in the following way:
\begin{itemize}
\item \textbf{Objective function:}
\begin{equation*}
\frac{|\mathbf{h_s}^\dagger\bm{\omega}|^2}{1+\bm{\omega}^\dagger\bm{\omega}}=|\mathbf{h_s^\dagger u}|^2, \text{ where } \mathbf{h_s}=[h_{s,1},\cdots,h_{s,M}]^\dagger
\end{equation*}

\item \textbf{Eavesdroppers Constraint} $\forall k \in \mathcal{K}$: 
\begin{align*}
\mathbf{h}_{s\rho,k}=[h_{s,1}\rho_{1,k},h_{s,2}\rho_{2,k}\cdots,h_{s,M}\rho_{M,k}]^\dagger\\   
\frac{\left| \sum_{i=1}^{M} h_{s,i} \beta_i h_{i,k}\right| ^2}{1 + \sum_{i=1}^{M}|\beta_ih_{i,k}|^2}=\frac{\bm{\omega}^\dagger(\mathbf{h}_{s\rho,k}\mathbf{h}_{s\rho,k}^\dagger)\bm{\omega}}{1+\bm{\omega}^\dagger \mathbf{D}_{\rho,k}\bm{\omega}}\le \gamma_k',
\end{align*}
where $\mathbf{D}_{\rho,k}=diag(\bm{\rho}_k)$ and $\gamma_k'=\gamma_k\frac{\sigma^2}{P_s},\;\forall k \in \mathcal{K}$. By writing in terms of $\mathbf{u}$ and rearranging, we get: $\mathbf{u}^\dagger\mathbf{C_ku} \le 1$, where 
\begin{equation*}
\mathbf{C_k}=\frac{\mathbf{h}_{s\rho,k}\mathbf{h}_{s\rho,k}^\dagger}{\gamma'_k}+\mathbf{I}-\mathbf{D}_{\rho,k}
\end{equation*}

\item \textbf{Relay power constraint} $\forall i \in \mathcal{M}$: 
\begin{align*}
& |\beta_i|^2=\frac{|\omega_i|^2}{|h_{i,d}|^2}=\frac{|u_i|^2}{|h_{i,d}|^2(1-\mathbf{u^\dagger u})} \le \beta_{i,max}^2\\
\text{ or }& \mathbf{u^\dagger u}+\frac{|u_i|^2}{|h_{i,d}|^2\beta_{i,max}^2} \le 1 \text{ or } \mathbf{u^\dagger D_iu} \le 1,
\end{align*}
where 
\begin{align*}
(\mathbf{D_i})_{jk}=
      \begin{cases}
      & 1+ \frac{1}{|h_{i,d}|^2\beta_{i,max}^2},\;\text{if }k=j=i\\
      & 1 ,\text{ if }k=j\ne i\\
      & 0, \text{ otherwise } 
      \end{cases}
\end{align*}
\end{itemize}
Therefore, the equivalent optimization problem is:
\begin{subequations}
\label{eq:snr_cvx}
\begin{align}
\text{\textbf{P1-EQ:}}\;\max_\mathbf{u}&\; \mathbf{u^\dagger h_sh_s^\dagger u}\\
\text{such that}&\;\mathbf{u^\dagger C_ku} \le 1,\;k \in \mathcal{K}\\
&\mathbf{u^\dagger D_iu} \le 1, \; i \in \mathcal{M}
\end{align}
\end{subequations}
If $\rho_{i,k}=\frac{|h_{i,k}|}{|h_{i,d}|}\le 1,\;\forall i,k$ then $\mathbf{I}-\mathbf{D}_{\rho,k}$ is a diagonal matrix with positive entries, therefore, $\mathbf{C}_k$ is a positive definite matrix.
As the objective function and all the constraints are convex, 
we can obtain the global optimum using numerical routines as suggested by \cite{hiriart2001global}.
\par But, in general scenarios $\mathbf{C}_k$ may not be positive-semidefinite.  However, if a \textbf{QCQP} is not convex then it is hard to solve \cite{ANDREA},\cite{luo2010semidefinite}. As $\mathbf{u^\dagger h_sh_s^\dagger u}=\text{Trace}(\mathbf{u^\dagger h_sh_s^\dagger u})=\text{Trace}(\mathbf{h_sh_s^\dagger uu}^\dagger)$, so by considering relaxation on rank one symmetric positive semi-definite (PSD) matrix $\mathbf{U}=\mathbf{uu}^\dagger$, the optimization program \ref{eq:snr_cvx} can be written as:
\begin{subequations}\label{eq:snr_sdr}
\begin{align}
\max_{\mathbf{U}}& \quad \mathbf{h_sh_s^\dagger \bullet U}\\
\text{subject to }& \quad \mathbf{C_k\bullet U} \le 1, k \in \mathcal{K}\\
& \quad \mathbf{D_i\bullet U} \le 1, i \in \mathcal{M}
\end{align}
\end{subequations}
The $\bullet$ represents Frobenius product of matrices, $\mathbf{A \bullet B}=\text{ Trace }(\mathbf{A^\dagger B})$. Once the problem \ref{eq:snr_sdr} is solved, we can find the corresponding optimal $\mathbf{u}$ and thereby $\mathbf{w}$ by applying eigenvalue decomposition on matrix $\mathbf{U}$ \cite{luo2010semidefinite}.

\textit{Remark 1:} If we denote the optimum objective function value by $SNR_d^*$, then source can securely transmit at a rate $\frac{1}{2}\log(1+\frac{SNR_d^*}{SNR_d^*-\gamma_d}[SNR_d^*-\gamma_d]^+)$,
 where $[x]^+=\max\{x,0\}$ and $\gamma_d$ is the SNR threshold for destination.
   
\section{Relay Power Minimization}
\label{POW}
   For an energy constrained relay network, an objective would be to minimize the total relay power. Using \eqref{eq:beta}, the power of the signal transmitted by the $i^{th}$ relay node is $|\beta_i|^2 (|h_{s,i}|^2 P_s + \sigma^2),\;\forall i \in \mathcal{M}$. Considering the SNR constraints on the destination and the eavesdroppers and individual power constraint for each relay node, the total relay power minimization problem is as follows:
   \begin{subequations}\label{prob1}
   \begin{align}
   \text{\textbf{ P2:} }\; \min_{\bm{\beta}} & \sum\limits_{i=1}^{M}|\beta_i|^2 (|h_{s,i}|^2 P_s + \sigma^2)\\
   \text{such that}\; & \frac{\left| \sum_{i=1}^{M} h_{s,i}\beta_ih_{i,d}\right| ^2}{1 + \sum_{i=1}^{M}|\beta_i h_{i,d}|^2}\frac{P_s}{\sigma^2} \ge \gamma_d \label{prob1cons1}  \\
   		& \frac{\left| \sum_{i=1}^{M} h_{s,i} \beta_i h_{i,k}\right| ^2}{1 + \sum_{i=1}^{M}|\beta_ih_{i,k}|^2}\frac{P_s}{\sigma^2} \le \gamma_k, \; k \in \mathcal{K} \label{prob1cons2}\\
    		&|\beta_i|^2 \le \beta_{max,i}^2, i \in \mathcal{M}
   \end{align}
   \end{subequations}
   Similar to $\gamma_k$, $\gamma_d$ is a real number and it represents the predefined SNR threshold for the destination.
   We can easily rewrite the above expression in matrix notation and rearrange to get the following Quadratic Constrained Quadratic Program (QCQP) formulation \cite{ANDREA}:
   \begin{align*}
   \text{\textbf{ P2-QCQP:} } 
   \;\min_{\bm{\beta}} \quad & \boldsymbol{\beta}^\dagger\mathbf{Q}_0\boldsymbol{\beta} \\
   & \boldsymbol{\beta}^\dagger\mathbf{Q}_D\boldsymbol{\beta} \ge \gamma'_d \\
   & \boldsymbol{\beta}^\dagger\mathbf{Q}_{k}\boldsymbol{\beta} \le \gamma'_{k} , \quad k \in \mathcal{K} \\
   & |\beta_i|^2 \le \beta_{max,i}^2, \; i \in \mathcal{M}
   \end{align*}
   \begin{align*}
   \text{ where }&\gamma'_l = \frac{\sigma^2}{P_s}\gamma_l,\quad l \in \{D\}\bigcup\mathcal{K}\\
   \mathbf{Q}_0= &\text{diag}(|h_{s,1}|^2P_s+\sigma^2,\;|h_{s,2}|^2P_s+\sigma^2
    \;\dots ,|h_{s,M}|^2P_s+\sigma^2)\\
   \mathbf{Q}_l= &\mathbf{h_{s,l}h_{s,l}}^\dagger-\gamma'_l\mathbf{D}_l,\;l \in \{D\} \bigcup \mathcal{K}
   \end{align*}
     Following the procedures discussed in previous section, we formulate another semi-definite program after relaxing the 
      symmetric positive semi-definite (PSD) matrix $\mathbf{B}=\bm{\beta\beta}^\dagger$.
\begin{subequations}
\label{prob:P2SDP}            
\begin{align}
\text{\textbf{P2-SDR:}}\;\min_\mathbf{B} \quad &  {\mathbf{Q}}_0 \bullet {\mathbf{B}} \\
    -& {\mathbf{Q}}_D \bullet {\mathbf{B}} \le -\gamma'_d\\
    & {\mathbf{Q}}_k \bullet {\mathbf{B}} \le \gamma'_k,\quad k \in  \mathcal{K}\\
    &\mathbf{B}_i \le \beta_{max,i}^2, \; i \in \mathcal{M}\\ 
    & {\mathbf{B}} \succeq 0
\end{align} 
\end{subequations}
 $\mathbf{B}_i$ represents the $i^{th}$ diagonal element of that matrix. Solving this will provide us $\mathbf{B}$; we look for an eigenvector of $\mathbf{B}$ corresponding to the maximum eigenvalue. 

All the above formulations of Problem P2 require iterative algorithms to obtain the optimal solution. Therefore, we discuss a simple analytical relaxed solution for the problem \eqref{prob1}.

\textit{\textbf{Analytical Relaxed Solution:}} It can be argued that to minimize the objective function value, \eqref{prob1cons1} should be satisfied with equality. Therefore, we formulate another optimization problem to calculate the analytical relaxed solution of the original objective function value.
\begin{subequations}
\begin{align}
\min_{\bm{\beta}}\quad& \bm{\beta}^\dagger\mathbf{Q}_0\bm{\beta} \\
\text{such that}\quad & \bm{\beta}^\dagger\mathbf{Q}_D\bm{\beta}=\gamma'_d
\end{align} 
\end{subequations}
This can be posed as the well known \textit{Rayleigh Quotient} \cite[p.~176]{Horn1985} problem \textit{i.e.}
\begin{equation}
\label{eq:rayleigh}
\min_{\bm{\beta}}\quad  \gamma'_d\frac{\bm{\beta}^\dagger\mathbf{Q}_0\bm{\beta}}{\bm{\beta}^\dagger\mathbf{Q}_D\bm{\beta}} 
\end{equation}
If $\mathbf{Q}_D$ is invertible, then the solution to this optimization problem is given by $\gamma'_d\lambda_{min}$, where $\lambda_{min}$ is the minimum eigenvalue of matrix $\mathbf{Q}^{-1}_D\mathbf{Q}_0$ and the optimal $\bm{\beta}$ is the corresponding eigenvector. The solution of this problem need not satisfy the eavesdroppers' constraints, but the objective function value serves as lower bound for optimization problem \ref{prob:P2SDP} and indicate the extra amount of power required to met the secrecy constraints.

\textit{Remark 2:} A lower bound on the achievable AF secrecy rate for the optimum scaling vector would be $\frac{1}{2}\log(1+\gamma_d)$.

\section{Results}
\label{sec:res}
\begin{figure}[!t]
\centering
\includegraphics[scale=0.6]{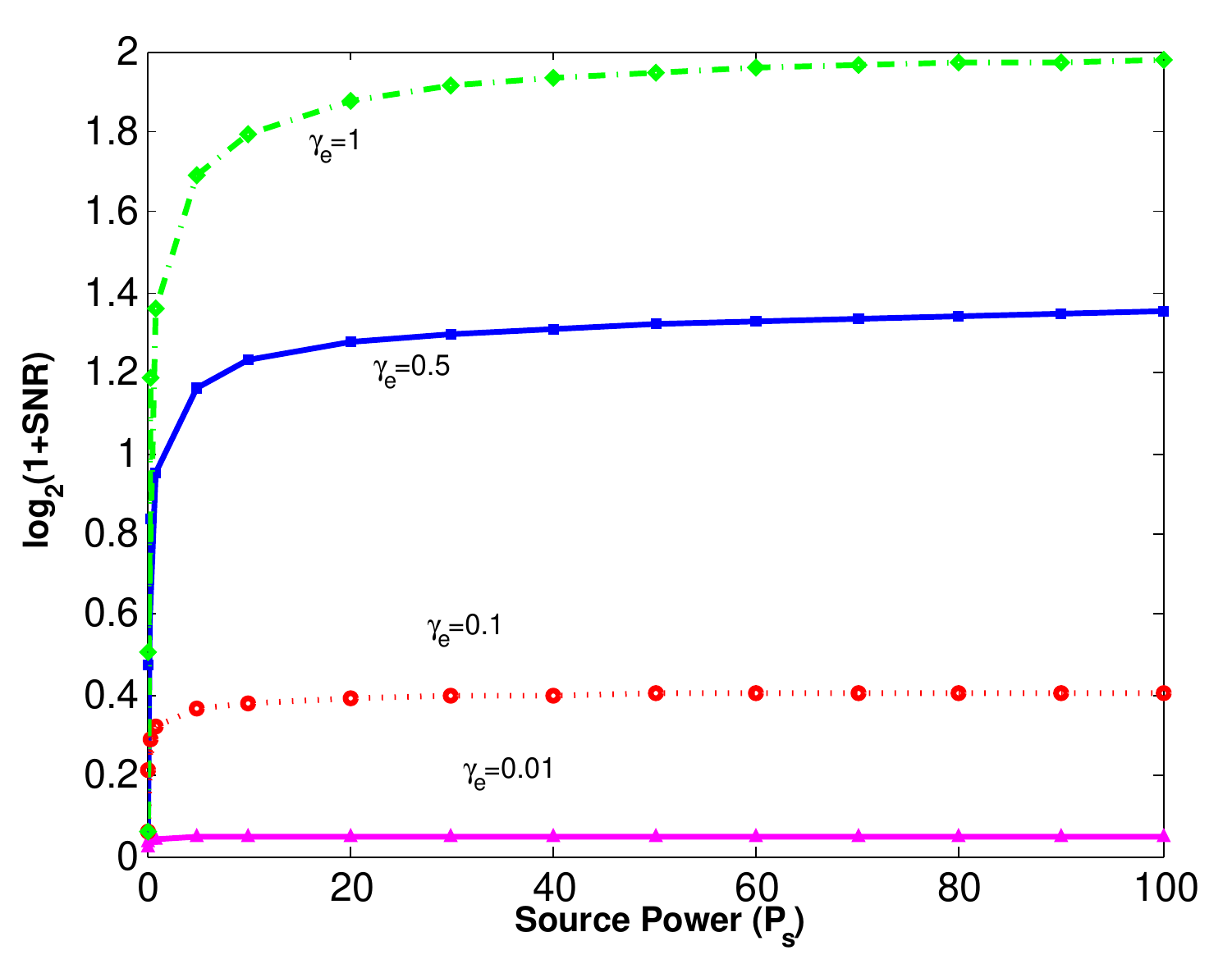}
\caption{Plot of the achievable secrecy rate at the destination $\log_2(1+SNR_d)$ obtained from SNR maximization problem (\textbf{P1}) w.r.t. to source power $(P_s)$ for several SNR threshold values $(\gamma_e)$ with $P_i=10,\;\sigma^2=1$, $M=5$ and $K=5$, averaged over 100 instances.}
\label{fig_snrmax}
\vspace{-0.2in}
\end{figure}

    In Figure \ref{fig_snrmax}, we plot the solution of the destination SNR maximization problem, mentioned in Section~\ref{SNR}, with respect to source power ($P_s$) for several values of eavesdroppers' SNR threshold ($\gamma_k=\gamma_e=\mbox{constant},\;k\in\mathcal{K}$) after averaging over 100 network instances. 
    For each instance, the channel gains are obtained from a complex Gaussian distribution with mean 0 and variance 1. We considered five relay nodes ($M=5$) with equal maximum transmit power ($P_i=10,\;i\in\mathcal{M}$) and five eavesdroppers ($K=5$) for the evaluation. Though we have assumed equal threshold values ($\gamma_e$) for all the eavesdroppers in our plots, our formulations are valid for unequal values ($\gamma_i \ne \gamma_j$ where $i\ne j, i,j \in \mathcal{K}$) also.
     As $\gamma_e$ increases, the relay nodes can transmit signals with higher power. This results in the increase of SNR at the destination, yet the received SNRs at the eavesdroppers remain below their respective SNR thresholds. This leads to an increase in the secure communication rate with increasing $\gamma_e$, which is evident from Figure \ref{fig_snrmax}.

\begin{figure}[!t]
\centering
\includegraphics[scale=0.6]{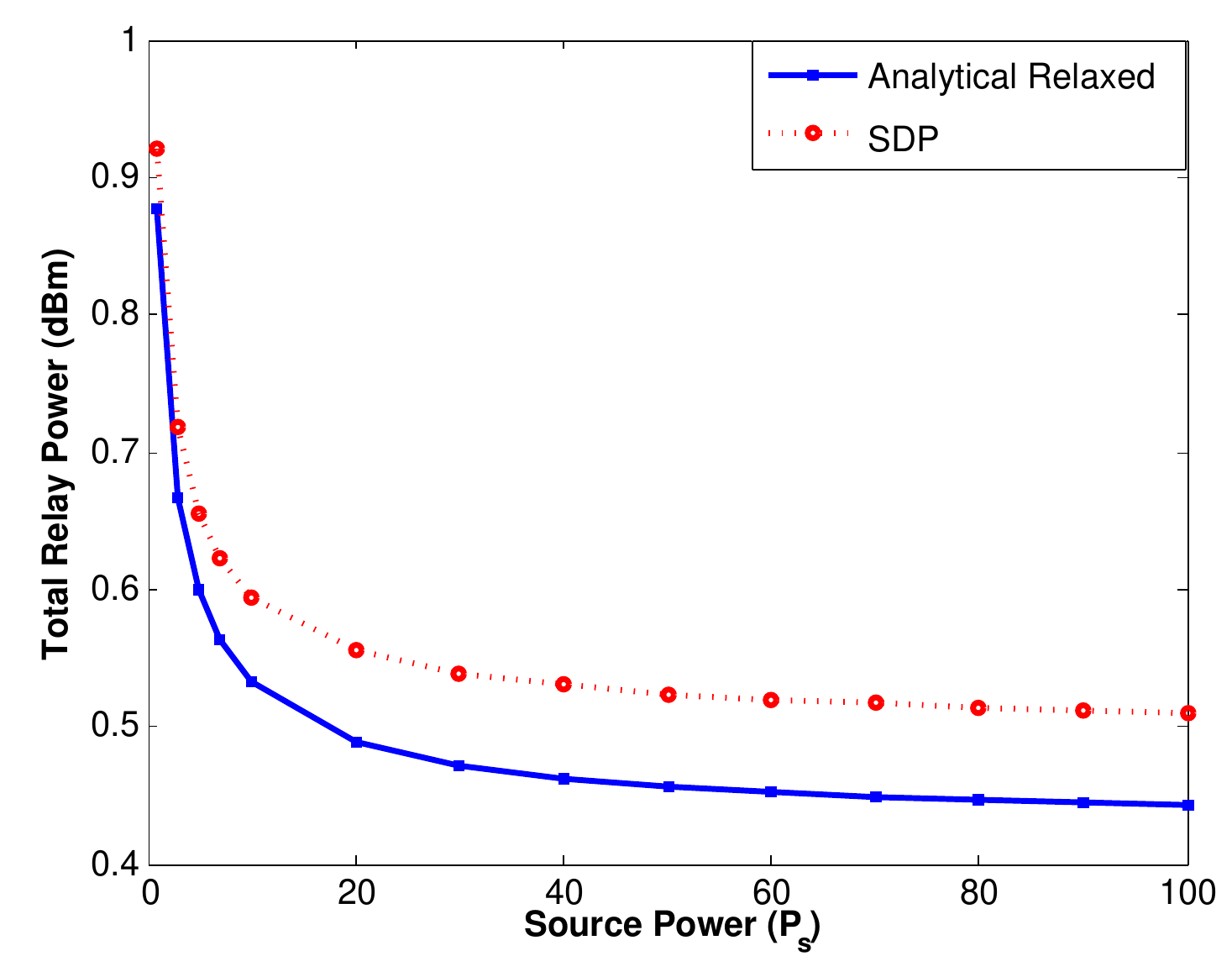}
\caption{Plot of total relay power consumption calculated using optimization formulations in \eqref{prob1},\eqref{prob:P2SDP}, and \eqref{eq:rayleigh} w.r.t source power ($P_s$) with $\gamma_d=0.01$ and $\gamma_e=0.005$, $P_i=10,\;\sigma^2=1$, $M=5$ and $K=5$, averaged over 100 instances.}
\label{fig_powmin}
\end{figure}

Figure \ref{fig_powmin} depicts the variation of optimal total relay power consumption for the total relay power minimization formulations, mentioned in Section~\ref{POW}, with respect to the source power ($P_s$).
      The ``SDR" and ``Analytical Relaxed" values are outputs of the relaxed semidefinite program \textbf{P2-SDR} and Rayleigh Quotient problem \eqref{eq:rayleigh}, respectively. 
      With the increase of source power $(P_s)$, the SNR at relay nodes increases. Therefore, to maintain the SNR at the destination at any fixed value, the relays have to consume less power. This is apparent from Figure \ref{fig_powmin}, which shows decreasing total relay power consumption with the increase of $P_s$. 
    
\section{Conclusion}
\label{sec:concl}
SNR based secure communication for Amplify-and-Forward (AF) relay networks with multiple eavesdroppers is considered. After motivating 
the secure communication scheme, we formulate two optimization problems for obtaining respective optimal scaling factors. Due to the non-convexity of both the formulations, we reformulate those for obtaining computationally efficient solutions. We have considered a two hop network, but in future we plan to investigate similar SNR based model for general multihop AF networks. We also plan to study scenarios where some relay nodes act as jammers, thereby reducing the SNR at the eavesdroppers \cite{sarma}, \cite{sankararaman2012}.

\end{document}